\documentclass[journal=jacsat,manuscript=article]{achemso}

\usepackage{chemformula} 
\usepackage[T1]{fontenc} 
\usepackage{xcolor}%
\usepackage{textcomp}%
\usepackage{manyfoot}%
\usepackage{booktabs}%
\usepackage{algorithm}%
\usepackage{algorithmicx}%
\usepackage{algpseudocode}%
\usepackage{listings}%
\usepackage{soul}
\usepackage{float}
\usepackage{lineno}
\usepackage{tabularx}
\usepackage[table,xcdraw]{xcolor}
\usepackage{colortbl}
\usepackage{natbib}

\author{Hangxu Liu}
\affiliation{SCNU Environmental Research Institute, Guangdong Provincial Key Laboratory of Chemical Pollution and Environmental Safety, School of Environment, South China Normal University, Guangzhou 510006, P. R. China.}
\alsoaffiliation{MOE Key Laboratory of Environmental Theoretical Chemistry, South China Normal University, Guangzhou 510006, P. R. China.}
\author{Yifei Zhu}
\affiliation{SCNU Environmental Research Institute, Guangdong Provincial Key Laboratory of Chemical Pollution and Environmental Safety, School of Environment, South China Normal University, Guangzhou 510006, P. R. China.}
\alsoaffiliation{MOE Key Laboratory of Environmental Theoretical Chemistry, South China Normal University, Guangzhou 510006, P. R. China.}
\author{Zhenggang Lan}
\affiliation{SCNU Environmental Research Institute, Guangdong Provincial Key Laboratory of Chemical Pollution and Environmental Safety, School of Environment, South China Normal University, Guangzhou 510006, P. R. China.}
\alsoaffiliation{MOE Key Laboratory of Environmental Theoretical Chemistry, South China Normal University, Guangzhou 510006, P. R. China.}
\email{zhenggang.lan@m.scnu.edu.cn}

\title
  {An Automated Framework for Analyzing Structural Evolution in On-the-fly Non-adiabatic Molecular Dynamics Using Autoencoder and Multiple Molecular Descriptors}

\begin{document}
\graphicspath{{Figures/}}








\newpage
\begin{abstract}
A major challenge in nonadiabatic molecular dynamics is to automatically and objectively identify the key reaction coordinates that drive molecules toward distinct excited-state decay channels. Traditional manual analyses are inefficient and rely heavily on expert intuition, creating a bottleneck for interpreting complex photochemical processes. To overcome this, we introduce a fully automated machine-learning framework that directly extracts these coordinates from on-the-fly trajectory surface hopping data. By combining an Autoencoder for nonlinear dimensionality reduction with clustering and information entropy analysis, our method autonomously maps reaction channels and pinpoints their governing structural motions. When applied to keto isocytosine and the methaniminium cation, the framework objectively revealed invovled reaction channels and corresponding active coordinates with high efficiency and accuracy. This work establishes an effective paradigm for mechanistic insight in excited-state dynamics, transforming raw trajectory data into clear, interpretable reaction mechanisms.
\end{abstract}

\newpage
\section{Introduction}

Nonadiabatic molecular dynamics (NAMD), a frontier area of physical chemistry, plays a critical role in photochemistry.~\cite{NAMD1,NAMD2,NAMD3,NAMD4} It mainly studies the electronic transitions driven by the nuclear motions in photoinduced molecular processes. 
In theoretical chemistry, the all-atom-level description of NAMD in realistic polyatomic systems is a formidable challenge, as it involves strongly coupled electronic and nuclear motions, numerous entangled degrees of freedom and large-amplitude molecular motions.
Over the past few decades, a plethora of NAMD simulation methods have been developed.~\cite{NAMD3,NAMD4,fullquantum2,fullquantum5,semiclassical2,Crespo,TSH}Among them, trajectory surface hopping (TSH)~\cite{Crespo,TSH,Granucci,WangLinjun,Akimov,Tapavicza,codeTSH} is widely used to study non-adiabatic processes. 
As a mixed quantum-classical approach, it includes three essential components: the classical description of the nuclear motions on a single electronic state with Newtonian equations, the quantum description of electronic evolution with time-dependent Schr\"odinger equations, and the description of molecular transition between different electronic states with stochastic hopping.
After the combination of the on-the-fly simulation, the TSH framework has received considerable attention in the simulation of nonadiabatic dynamics of realistic polyatomic systems, mainly due to its broad application in describing arbitrary molecular motions, its simplicity and ease of implementation.~\cite{semiclassical2,Crespo,Tapavicza,codeTSH,Nelson1,Nelson2}

In TSH dynamics simulations, in principle, statistical analysis of these trajectories can yield key dynamic characteristics, including excited-state lifetime, reaction channel branching ratio, and dominant molecular motions. Typically, a large number of trajectories need to be propagated until statistical convergence is reached. In addition, realistic systems involve numerous degrees of freedom, and many of these motions are highly coupled.
As a result, the TSH dynamics simulation may generate a large quantity of high-dimensional data that is not easy to analyze.~\cite{Tully,Zhu2024} Especially for some systems that exhibit multi-channel and multi-product behaviors, effective data analysis becomes extremely challenging.~\cite{Zhu2024}
Traditional analytical methods typically rely on manual operations, for instance directly comparing the geometric differences between hopping and initial states, visually tracking geometric evolution during dynamics, and plotting relevant internal coordinates over time. Effective analysis based on these traditional methods becomes increasingly difficult as the system size increases, molecular motions become more complex and entangled, and the number of trajectories increases.

Machine learning (ML) provides a novel approach to addressing these challenges,~\cite{Zhu2024,QuantumChemistryML,Westermayr,NMA2,NMA3,NMA4,NMA5,Tsutsumi,D5SC05579B} especially through the dimensionality reduction algorithms in unsupervised learning. The dimensionality reduction algorithms are used to alleviate the difficulty in analyzing and visualizing high-dimensional data.~\cite{PCA1,PCA2,MDS1,MDS2,diffusionmap1,diffusionmap2,ISOMAP1,ISOMAP2} Its role is to compress high-dimensional data into a low-dimensional space, and during this compression process, the core information of the original data is preserved to ensure no critical features are lost due to simplification.
Following this idea, several attempts have been made to apply dimensionality reduction approaches to analyzing the geometry evolution in nonadiabatic dynamics. As the simplest linear dimensionality reduction approach, the principal component analysis (PCA)~\cite{PCA1,PCA2} exhibits strong capability in analyzing NAMD results.~\cite{Capano,Peng,Zhu2022,Delmas,Rukin,Lijingbai} 
Another linear dimensionality reduction approach, multidimensional scaling (MDS),~\cite{MDS1,MDS2} was also adopted to NAMD analysis.~\cite{Lixusong1,Lixusong2,Linkunni,Taketsugu,8b00176,s41061}
In excited-state dynamics simulations, pronounced nonlinearities may arise due to highly mixed molecular motions. To describe situations where the data are distributed on a nonlinear manifold, the nonlinear dimensionality reduction algorithms, such as diffusion map~\cite{diffusionmap1, diffusionmap2, Virshup, Richings, Domcke} or isometric feature mapping (ISOMAP)~\cite{ISOMAP1, ISOMAP2, Lixusong1} were introduced.



However, it is quite likely that different conical intersections (CIs)-driven reaction channels appear in the nonadiabatic dynamics. Clearly, the direct application of the above protocol becomes formidable as different channels are governed by their own specific coordinates.
In these cases, the clustering algorithms of unsupervised ML can be utilized to identify group patterns.~\cite{Kmeans1,Kmeans2,Kmeans3,hierarchical1,hierarchical2,DBSCAN1,DBSCAN2,DBSCAN3,DONKEY,CATBOSS} The core of clustering is grouping similar data, which automatically categorizes similar data points in the corresponding low-dimensional space into the same cluster. The K-means,~\cite{Kmeans1,Kmeans2,Kmeans3} hierarchical clustering~\cite{hierarchical1,hierarchical2} and the density-based spatial clustering of applications with noise (DBSCAN) approach~\cite{DBSCAN1,DBSCAN2,DBSCAN3} and other algorithm~\cite{DONKEY,CATBOSS} were applied in this aspect.~\cite{Linkunni,Kochman,Zhu2022,Lixusong2,Acheson}

These available works have already provided promising results demonstrating the effectiveness of unsupervised ML methods in the automated analysis of nonadiabatic dynamics. 
In additional,  ULaMDyn,~\cite{ULaMDyn} an open-source Python package was designed to automate the unsupervised learning analysis of large NAMD datasets.
However, further exploration of several key elements in this type of analysis protocol is still necessary to facilitate their broad application.

Firstly, how to select the proper descriptors?
Molecular descriptors, which correspond to high-dimensional vectors, are designed to adequately represent molecular structures. Different descriptors, such as the root mean squared error (RMSD) matrix,~\cite{Lixusong1,Lixusong2,Virshup} Coulomb matrix,~\cite{CM1,CM2,CM3,CM4} inverse distance matrix (IDM),~\cite{IDM1,IDM2} internal coordinates,~\cite{Zhu2022,Virshup,8b00176} etc., encode different geometric features of molecular structures.
This indicates that the ML analysis results may be strongly dependent on the descriptor selection, while studies on this dependence remain scarce.

Secondly, how to more efficiently and effectively identify different reaction channels for multi-channel problems? Although this issue can be tackled with clustering approaches, the analysis protocol requires an appropriate metric to classify trajectories into different reaction channels. For instance, by measuring the trajectory similarity by the the Fr\'{e}chet distance,~\cite{Fréchet1,Fréchet2,Fréchet3} the clustering analysis was performed in the trajectory space.~\cite{Lixusong2,Kochman,DONKEY,Acheson}
Alternatively, we may divide hopping geometries into different groups to identify reaction channels.~\cite{Zhu2022} To deal with the situations with the strong mixing of various reaction coordinates, the hierarchical and iterative algorithms have been proposed to classify these geometries into different groups according to their geometric similarity. However, all these protocols involve rather complex and cumbersome steps, largely preventing their broad application.


Thirdly, how to clarify the active reaction coordinate for each channel? After identifying reaction channels, it is also crucial to ascertain the reaction coordinates for each identified cluster.
However, this task is also far from straightforward, because many dimensionality reduction approaches do not provide direct transformation from the original high-dimensional coordinates to the reduced low-dimensional coordinates.




To address these challenges, we propose a highly automated framework to identify different reaction channels and their corresponding key molecular motions from on-the-fly TSH dynamics simulations.
This protocol consists of two steps: the first is to identify reaction channels, and the second is to identify the major active coordinates of each channel.

In the first step, we utilize the Autoencoder~\cite{Autoencoder1,Autoencoder2,DeepLearning} combined with a variety of molecular descriptors to reduce the dimension of molecular data. 
In fact, the Autoencoder methods have already been adopted to analysis ground state molecular dynamics.~\cite{Regina1,Regina2}
Additionally the performance of different descriptors is examined according to the results.
In the reduced space, the DBSCAN algorithm~\cite{DBSCAN1,DBSCAN2,DBSCAN3} is employed to divide all data points into several clusters, in principle, each cluster represents a reaction channel.

In the second step, to simplify the direct characterization of reaction channels and obtain a complete analysis framework, we compare the hopping geometries with their corresponding initial geometries using information entropy.~\cite{Entropy1,Entropy2,Entropy3,Entropy4}
In fact, the information entropy was once utilized to analyze the nonadiabatic dynamics outputs.~\cite{Tavadze,Prezhdo1,Prezhdo2,Prezhdo3,Prezhdo4}. 
We can utilize information entropy to quantify the key driving parameters during molecular changes. After seamless integration of the aforementioned steps, we developed a highly automated protocol to analyze the nonadiabatic dynamics simulation results efficiently and effectively.

The excellent performance of this analysis protocol was validated with two typical examples, the keto isocytosine~\cite{Nakayama,Mai,Gonzalez,Keto1,Keto2,Zhu2022,C6CP01391K} and methaniminium cation.~\cite{CH4N1,Lixusong1,Newton-X1,Fabiano,codeTSH,Tapavicza,Barbatti}
Overall, this work should facilitate the establishment of a black-box, easy-to-use analysis protocol applicable to more general types of nonadiabatic dynamics simulations.




\newpage
\section{Methods}

We developed a comprehensive computational framework for automatically analyzing simulation results from TSH dynamics, which can be readily extended to other trajectory-based nonadiabatic dynamics approaches. This protocol integrates several key components: TSH dynamics simulations, molecular descriptor construction, Autoencoder-based dimensionality reduction, clustering analysis, and information entropy theory.

\subsection{TSH Dynamics}

TSH~\cite{Crespo,TSH,Granucci,WangLinjun,Akimov,Tapavicza,codeTSH} dynamics represents a mixed quantum-classical methodology. Within this framework, classical mechanics describes nuclear motion on a single potential energy surface, while electronic evolution is captured by solving the time-dependent Schr$\ddot{o}$dinger equation. To model non-adiabatic transitions, the TSH algorithm allows instantaneous hops between different electronic states, with transition probabilities determined by Tully's fewest-switches surface hopping scheme~\cite{TSH} combined with the energy-based decoherence correction.~\cite{Granucci}
For convergence a swarm of trajectories are calculated to obtain the averaged physical quantities in the nonadiabatic dynamics.

\subsection{Descriptors}

Analyzing molecular evolution in nonadiabatic dynamics requires appropriate representations to characterize molecular geometries.~\cite{Zhu2024,QuantumChemistryML}
Molecular descriptors, encoded as high-dimensional vectors or tensors, serve this purpose. 
Normally molecular descriptors are categorized into global and local types. The former one characterizes the overall molecular properties, while the latter one emphasizes each specific region of the molecule.

Six popular molecular descriptors are chosen in this study, as they emphasize different molecular geometry features. The descriptors include: Cartesian coordinates, redundant internal coordinates (RIC),~\cite{RIC1,RIC2} IDM, many-body tensor representation (MBTR),~\cite{MBTR} smooth overlap of atomic positions (SOAP)~\cite{SOAP} and atomic environment vectors (AEV).~\cite{AEV} 
More detailed discussions on the implementation are given in the below sections and in SI.

\subsubsection{Cartesian Coordinates}

Cartesian coordinates provide a straightforward representation of molecular structures, with each atom's position defined by its (x, y, z) values in Euclidean space. As these coordinates are naturally employed in nonadiabatic dynamics simulations, they represent a logical choice for characterizing geometric evolution. However, to focus exclusively on internal molecular motions, the influences of translational and rotational motions must be eliminated. Accordingly, we utilize ``aligne'' Cartesian coordinates constructed by superimposing geometries using the Kabsch algorithm.~\cite{Kabsch1,Kabsch2} For chiral systems, appropriate reflection operations are also considered to account for mirror image effects.

\subsubsection{Redundant Internal Coordinates (RICs)}

Internal coordinates offer a chemically intuitive representation of molecular geometries by automatically incorporating connectivity information through bond lengths, bond angles, and dihedral angles. A limitation of this approach is the non-uniqueness in defining internal coordinates, which complicates appropriate selection. We employ RICs~\cite{RIC1,RIC2} due to their comprehensive representation of molecular connectivity, following the standard implementation in Gaussian software.~\cite{codeg16} The construction principles include: (1) identifying chemical bonds between Atoms A and B based on interatomic distances with respect to predefined criteria; (2) defining bond angles for three sequentially bonded atoms (A-B-C); and (3) determining dihedral angles for four sequentially bonded atoms (A-B-C-D). The resulting RIC set comprises all valid bond lengths, bond angles, and dihedral angles defined through this procedure.

\subsubsection{Inversed Distance Matrix (IDM)}

The IDM is constructed by computing the inverse values ($1/R_{ij}$) of all pairwise atomic distances ($R_{ij}$). 
Similar to the Coulomb matrix,~\cite{CM1,CM2,CM3,CM4} it serves as an effective global molecular descriptor that emphasizes the three-dimensional geometric features.
For a molecule containing N atoms, the vector derived from the IDM contains $N(N-1)/2$ features due to the symmetry of the distance matrix.

\subsubsection{Many-Body Tensor Representation (MBTR)}

The MBTR~\cite{MBTR} is a global molecular descriptor widely used in ML models to predict molecular properties such as energy and force. 
For a molecule with $N$ atoms of atomic numbers $z_i$ and positions $R_i$, MBTR encodes structural information by systematically considering correlations among one, two, and three atoms ($k = 1, 2, 3$). 
The general form is
\begin{equation}
    f(k, x, Z) = \sum_{\mathbf{i}} w(k,\mathbf{i})\, D[x, g(k,\mathbf{i})] 
    \prod_j^k \delta(z_{\mathbf{i}_j}, Z_j),
\end{equation}
where $\mathbf{i} = \{i_1, \ldots, i_k\}$ denotes a unique $k$-tuple of atoms, $\delta$ is Kronecker’s delta ensuring atom-type matching, and  $Z=\{Z_1,\ldots,Z_k \}$ is the atomic space designated of the $k$ atoms.

The geometric function $g(k,\mathbf{i})$ transforms each atomic group into a scalar quantity:
\begin{equation}
g(1) = z_{i_1}, \quad 
g(2) = |R_{i_1}-R_{i_2}|, \quad 
g(3) = \angle(R_{i_1},R_{i_2},R_{i_3}),
\end{equation}
representing atomic number, interatomic distance, and bond angle (or their variants such as inverse distance or cosine of angle), respectively.


Each scalar value $x$ is broadened by a Gaussian kernel with standard deviation $\sigma$, as
\begin{equation}
D[x,g] = \frac{1}{\sigma \sqrt{2\pi}} 
\exp\!\left[-\frac{(x-g)^2}{2\sigma^2}\right],
\end{equation}
to produce a continuous distribution.
The weighting term $w(k,\mathbf{i})$ decays exponentially with interatomic distance, effectively introducing a distance cutoff. 
All $f(k, x, Z)$ terms are concatenated over $k$ and element combinations $Z$ to form the final MBTR vector describing the molecule.

\subsubsection{Smooth Overlap of Atomic Positions (SOAP)}

The SOAP~\cite{SOAP} is a well-known local molecular descriptor developed by Langer et al, which provides a robust description of the local atomic environment for each atom in a molecular system.  
The construction of SOAP follows the below procedure. 
The atomic neighborhood density function $\rho^Z(\mathbf{r})$ with atom species $Z$ is defined as the superposition of of Gaussian functions centered on neighboring atoms. 
\begin{equation}
    \rho^{Z}(\mathbf{r}) \equiv \sum_{i \in Z} \exp(-\alpha|\mathbf{r}-\mathbf{r}_i|^2) = \sum_{nlm} c_{nlm}^{Z}g_n(r)Y_{lm}(\theta, \phi),
\end{equation}
where $\alpha$ controls the smoothness of the density function, $\mathbf{r}_i$ is the position of each neighbor atom. 
Similar to atomic orbitals in quantum mechanics using spherical coordinate system $(r, \theta, \phi)$, 
$\rho^Z(\mathbf{r})$ is further decomposed into the product of a radial basis function and a spherical harmonic function. 
Here $g_n(r)$ is the radial basis function, by default using spherical Gaussian type orbitals for computational efficiency, 
and $Y_{lm}(\theta,\phi)$ is the real spherical harmonic function. 
The $n$, $l$, $m$ are quantum numbers of spherical Gaussian type orbitals, 
and the values of $g_n(r)$ and $Y_{lm}(\theta,\phi)$ are determined by the values of $n$, $l$, $m$. 
The coefficient $c^Z_{nlm}$ is obtained by:
\begin{equation}
    c_{nlm}^Z = \iiint_{R^3} g_n(r) Y_{lm}(\theta, \phi) \rho^Z(\mathbf{r}) dV.
\end{equation}

The atomic SOAP output is the partial power spectrum vector $p$, where the elements are defined as:
\begin{equation}
    \mathbf{p}_{nn'l}^{Z_1 Z_2} = \pi \sqrt{\frac{8}{2l+1}} \sum_m c_{nlm}^{Z_1} (c_{n'lm}^{Z_2})^*,
\end{equation}
where the $(c_{n'lm}^{Z_2})^*$ is the conjugate of $c_{n'lm}^{Z_2}$. The elements in $\mathbf{p}$ traverse all the combinations of $Z_1$, $Z_2$, $n$, $n'$, $l$ and avoid repetition. 
Then we concatenate all the SOAP of atoms in the molecule into a molecular descriptor.

\subsubsection{Atomic Environment Vectors (AEV)}

Based on Behler-Parrinello symmetry functions, the AEV~\cite{AEV} descriptor systematically encodes both radial and angular features of each atom's local chemical environment.
The construction of AEV follows the below procedure. Firstly, for a central atom, a continuously smooth truncated function $f_C(R_{ij})$ is introduced to focus on its local chemical environment, in which two-body interaction between atom $i$ and atom $j$ is considered only when their distance $R_{ij}$ is less than the pre-defined truncated radius $R_c$, the $f_C(R_{ij})$ is defined by

\begin{equation}
    f_c(R_{ij}) = 
    \begin{cases} 
    0.5 \times \cos \left( \frac{\pi R_{ij}}{R_c} \right) + 0.5 & \text{for } R_{ij} \leq R_c, \\ 
    0.0 & \text{for } R_{ij} > R_c.
    \end{cases}
\end{equation}

This function reflects the atomic local environments defined by two-body interactions. 
To enhance resolution across different atomic types, AEVs are grouped according to atomic species. 
Specifically, for a $N$-atom species molecule, 
$N$ radial symmetry function $G^R$ and $N(N+1)/2$ angular symmetry function $G^A$ are generated 
to describe the environmental features of each type atom, respectively. 
Here $G^R$ are given as follows:
\begin{equation}
    G^R = \sum_{j \neq i} \exp[-\eta(R_{ij}-R_s)^2] \cdot f_C(R_{ij}),
\end{equation}
where the $R_s$ and $\eta$ are used to control the center and width of the radial Gaussian distribution, respectively.
And  $G^A$ are defined as:
\begin{equation}
    G^A = 2^{1-\zeta} \sum_{j,k \neq i} (1 + \cos(\theta_{ijk} - \theta_s))^\zeta \times \exp[-\eta(\frac{R_{ij} + R_{ik}}{2} - R_s)^2] \cdot f_C(R_{ij}) \cdot f_C(R_{ik}),
\end{equation}
where $\theta_s$ and $\zeta$ control the center and width of the angular Gaussian distribution respectively. 
By varying $R_s$ and $\theta_s$ for different combinations of atomic types, the AEV of the center atom is obtained, Finally  the AEVs of all atoms are concatenated to give a completed molecular descriptor.

\subsection{Autoencoder}

Dimensionality reduction techniques constitute a family of unsupervised machine learning methods that transform original high-dimensional datasets into reduced low-dimensional representations while minimizing information loss about the underlying data distribution patterns. 
This study employs the Autoencoder~\cite{Autoencoder1,Autoencoder2,DeepLearning} which based on neural network architecture due to its inherent advantages, particularly its ability to provide reasonable dimensionality reduction for both linear and nonlinear data. 
This flexibility enables the direct utilization of diverse molecular descriptors as input vectors for the Autoencoder.

\begin{figure}[ht]
    \centering
    \includegraphics[width=0.9\linewidth]{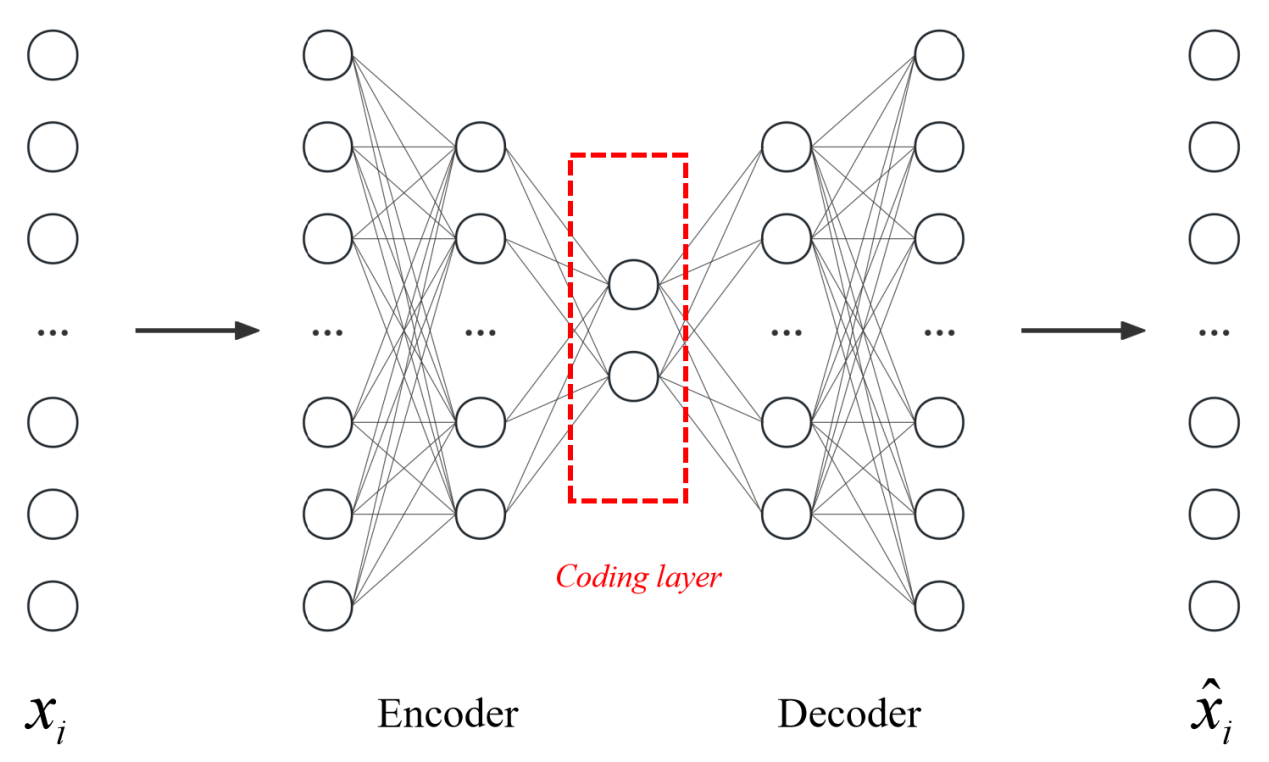}
    \caption{
    Schematic illustrations of the Autoencoder architecture.
    }
    \label{fig:fig1}
\end{figure}

The Autoencoder neural network comprises two key components: an Encoder and a Decoder, as schematically illustrated in Figure~\ref{fig:fig1}. The Encoder maps input data (molecular descriptors) into a low-dimensional representation within the latent space, while the Decoder attempts to reconstruct the original input data from these latent representations. During model training, the loss function quantifies the discrepancy between the original input data and the data reconstructed by the Decoder. Through this design, the coding layer naturally emerges as a low-dimensional latent representation, effectively serving as the result of dimensionality reduction.

\subsection{Clustering}

Clustering represents another category of unsupervised machine learning techniques, 
designed to partition all data points into distinct groups while maximizing intra-cluster similarity and minimizing inter-cluster dissimilarity. 
Applying clustering algorithms enables the grouping of different molecular structures into several clusters, 
with the fundamental expectation that all structures within the same cluster share significant geometric similarities.

This study employs DBSCAN~\cite{DBSCAN1,DBSCAN2,DBSCAN3} to identify regions of high density and low density within the data distribution, 
where each high-density area forms a distinct cluster, 
while low-density regions define cluster boundaries. 
The DBSCAN method relies on two key parameters, i.e.\ $\varepsilon$ (the neighborhood radius) and $\mathrm{MinPts}$ (the minimum number of samples required to form a cluster). For any data point, all data points whose distance from it is less than $\varepsilon$ are called its $\varepsilon$-neighborhoods, and if the number of its $\varepsilon$-neighborhoods is greater than $\mathrm{MinPts}$, it is called the core point. A core point and its $\varepsilon$-neighborhoods belong to the same cluster. Points not belonging to any cluster are labeled as noise.

 \subsection{Information Entropy}

Although the Autoencoder provides a reasonable reduced space spanned by latent vectors, establishing direct relationships between these latent vectors and the original molecular coordinates remains challenging. The identification of key reactive motions is achieved by establishing correspondence between molecular descriptors and reduced coordinates within the framework of information entropy theory.~\cite{Entropy1,Entropy2,Entropy3,Entropy4}

Information entropy represents a fundamental concept in information theory, quantifying the richness and redundancy of information. While multiple definitions of information entropy exist, this work employs conditional entropy, which describes the information uncertainty of a target variable  ($\mathbf{Y}$)  under the influence of a predictor variable ($\mathbf{X}$)—in this context.
In one word, we used the conditional entropy to build the correlation between the the feature $\mathbf{X}$ and the clusters $\mathbf{Y}$ in the current work.

The conditional entropy is defined as follows. For a variable set $\mathbf{X} = {x_1, x_2, \dots, x_n}$ and a target set $\mathbf{Y} = {y_1, y_2, \dots, y_m}$, the conditional entropy $H(\mathbf{Y}|\mathbf{X})$ is given by
\begin{equation}
    H(Y|X) = -\sum^n_{i} p(x_i) \sum^m_{j} p(y_j|x_i) \log_2 [p(y_j|x_i)],
\end{equation}
where $p(x_i)$ is the probability of $x = x_i$ and $p(y_j|x_i)$ is the probability of $y = y_j$ under the condition of $x = x_i$.
More intuitively, the conditional entropy quantifies the remaining uncertainty of one random variable ($\mathbf{Y}$) when another variable ($\mathbf{X}$) is known.
Higher uncertainty corresponds to larger entropy values.

In this study,  we aim to identify  which molecular motions contribute significantly to the reactive coordinates,
\textit{i.e.} finding the important features that assign all geometries into different clusters.
This task is accomplished through conditional entropy analysis, 
in which the clustering results are defined as the target variables $\mathbf{Y}$, the molecular descriptors are defined as the variables $\mathbf{X}$, and their conditional entropy is calculated.
Normally the smaller conditional entropy value indicates the stronger influence of the feature ($\mathbf{X}$) on the target variable ($\mathbf{Y}$).
This approach elucidates the dependence of the clustering analyses ($\mathbf{Y}$) on molecular geometry descriptors ($\mathbf{X}$).

In implementation, we collect all hopping geometries belonging to this signle decay channel and compare them with their corresponding initial structures. Here, RIC is chosen as the typical descriptor to illustrate the above protocol based on the information entropy in detail.
This selection is rational due to the below reasons: first RIC really gives the excellent performances in the below two example; second other descriptors may be treated in the similar way. 
In fact, other descriptors are equally capable of performing this task, as long as each feature within these descriptors possesses a well-defined physical interpretation. The opposite is true for spectral descriptors, like MBTR, which cannot do this.


Each elements of RIC are taken to calculate the discrete $\mathbf{X}$ variables in the information entropy analyses.
Here all RIC values were standardized within the $[0,1]$ interval using min-max scaling protocol, which is further partitioned into 10 value-based bins, labelled as 0 to 9 respectively.  This procedure  defines all $X_i$ variables. For instance, the values within the $[0,0.1)$ interval will be marked as $X_i=0$, and and so forth, with values in [0.9, 1] assigned $X_i=9$. 
For each decay channel, we used the binary values for $\mathbf{Y}=0 \, \,  \mbox{or} \, \,   1 $ to  distinguish the initial geometries ($\mathbf{Y}=0$) or hopping geometries ($\mathbf{Y}=1$) in the conditional entropy analysis. 
As both $\mathbf{X}$ and $\mathbf{Y}$ are defined, we compute their conditional entropy. 

This processing method is applicable to all descriptors with continuous numerical descriptors, while for discrete numerical descriptors such as molecular fingerprints, the original values can be directly used as $X_i$.



\subsection{Implementation}

\begin{scheme}
    \centering
    \includegraphics[scale=0.7]{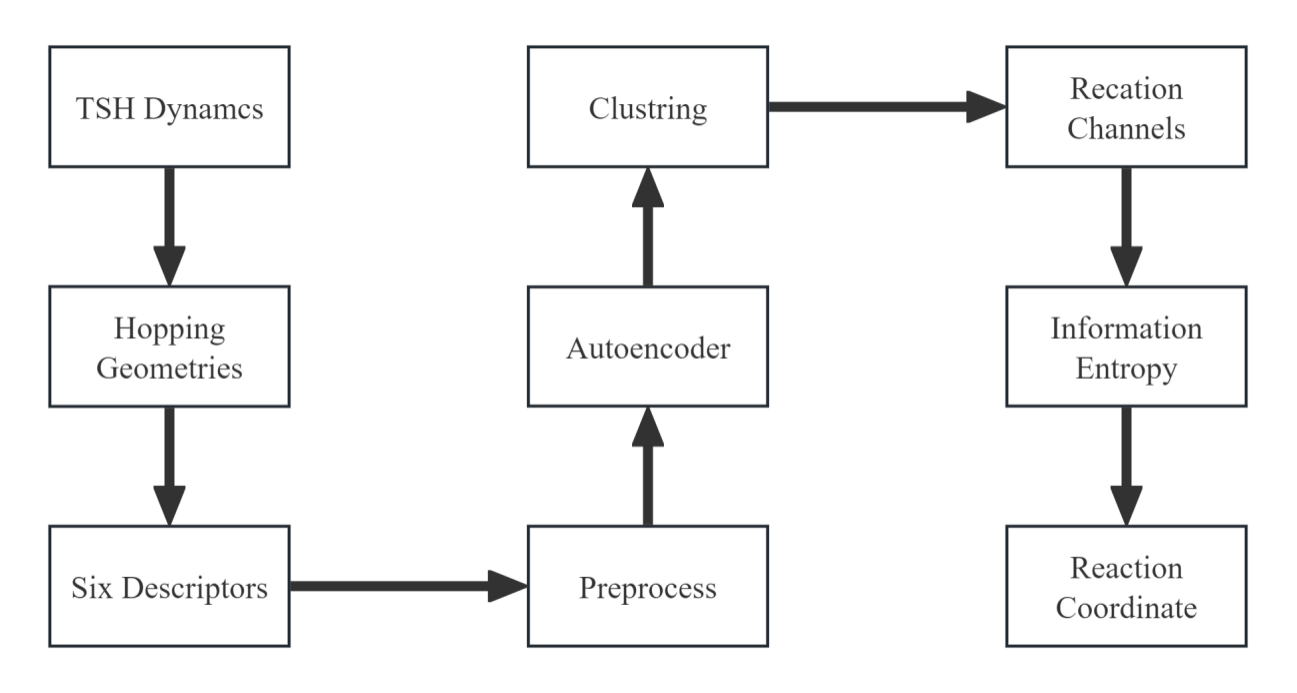}
    \caption{The workflow chart of the current analysis framework..}
    \label{flow_chart}
\end{scheme}

We now present the step-by-step working protocol with key implementation details. 
The overall workflow is illustrated in Scheme 1, with additional details available in the SI.

\begin{enumerate}
\item \textbf{TSH dynamics Simulations.} 
Normally, the TSH dynamics simulation includes the following steps. 
First, A set of initial nuclear configurations and velocities are generated. 
Second, the on-the-fly TSH dynamics simulation is performed and numerous trajectories are taken to compute the average values of physical observable, such as electronic populations. 
Based on the TSH simulation results, we initiate the analysis process.

\item \textbf{Data collection.} 
All hopping structures are collected, which are represented using six molecular descriptors mentioned above, \textit{i.e.} Cartesian coordinates, RIC, IDM, MBTR, SOAP and AEV. As the outlier data are typically treated as noise, we removed them in the further analyses through the following procedure. The molecular connectivity of each structure is examined according to the RIC definition. When the number of geometries with specific molecular connectivity is less than 1\% of the dataset, we treat them as outlier data and remove them in the further analysis to ensure statistical robustness. 
Additionally, considering mirror symmetry properties, we reverse the z-coordinate signs of selected geometries to eliminate chiral effects.

\item \textbf{Data Preprocessing.} 
Additional preprocessing procedures handle the diverse molecular descriptors. 
For Cartesian coordinates, IDM, and RIC~\cite{RIC1,RIC2} data, we apply min-max scaling to each feature independently. 
For the MBTR,~\cite{MBTR} SOAP,~\cite{SOAP} and AEV~\cite{AEV} descriptors, which possess extremely high dimensions, we first perform dimensionality reduction using PCA~\cite{PCA1,PCA2} to eliminate linear dependencies among features. 
All significant principal components (cumulative variance contribution >0.99) are retained as effective features, followed by a unified scaling approach for data normalization.

\item \textbf{Dimensionality Reduction.} 
Following preprocessing, all effective descriptors are processed through an Autoencoder neural network architecture.~\cite{Autoencoder1,Autoencoder2,DeepLearning}
The Autoencoder compresses the input data into a two-dimensional latent representation for visualization purposes. 
While data can be transformed to latent spaces with slightly higher dimensionality, we primarily consider two dimensions to maintain transparent visualization.

\item \textbf{Clustering Analysis.} 
The dimensionality reduction outputs are subjected to the clustering analysis with the DBCAN algorithm.~\cite{DBSCAN1,DBSCAN2,DBSCAN3}
In principle, each cluster should correspond to a distinct nonadiabatic decay channel. 

\item \textbf{Reaction Coordinate Identification.} 
To evaluate the contribution of individual input features in the clustering analyses, the information entropy (specially conditional entropy)~\cite{Entropy1,Entropy2,Entropy3,Entropy4} is adopted to identify the key coordinates that character 
the difference between the initial and hopping geometries, and providing the main features of the geometry evolution in each  channel.

\end{enumerate}

In the current work, the TSH dynamics were simulated using our long-term developing JADE code~\cite{codeTSH} interfaced with the MOLPRO package.~\cite{codeMOLPRO} All analysis codes were written by Python language, in which different functions are realized by calling to different libraries.
In the construction of molecular descriptors, the RIC are constructed using Gaussian 16 program package,~\cite{codeg16} the MBTR and SOAP are built based on DScribe, ~\cite{codeDScribe1,codeDScribe2} and and the AEV are generated based on TorchANI.~\cite{codeAEV} In the data preprocessing and clustering, the PCA and DBSCAN methods are 
realized based on scikit-learn.~\cite{codeSKlearn} In the dimensionality reduction step, the Autoencoder is written based on PyTorch.~\cite{codepytorch1,codepytorch2} The necessary codes are also available on GitHub (https://github.com/liuhx6368/Autoencoder.git)

\begin{figure}[ht]
    \centering
    \includegraphics[width=0.9\linewidth]{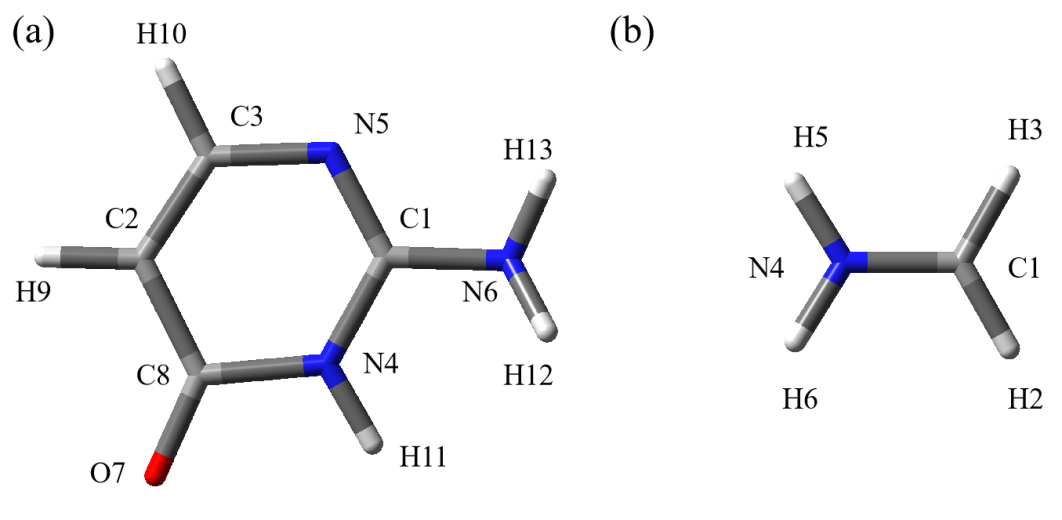}
    \caption{
    The S0 minimum structures of (a) keto isocytosine and (b) methaniminium cation.
    }
    \label{fig:fig2}
\end{figure}

In this work, two molecular systems, keto isocytosine~\cite{Nakayama,Mai,Gonzalez,Keto1,Keto2,Zhu2022,C6CP01391K} and methaniminium cation,~\cite{CH4N1,Lixusong1,Newton-X1,Fabiano,codeTSH,Tapavicza,Barbatti} were chosen as prototypes to explain the current analysis approach. Their $S_0$ minimum structures are shown in the Figure~\ref{fig:fig2}. As both were widely studied in many previous works, we did not provide a very comprehensive discussion on their nonadiabatic dynamics. Instead we focused on how to employ the current analysis protocol to identify the reaction channels and corresponding reaction coordinates.

Both the two model systems, the trajectories start from $S_2$. More details on the TSH dynamics and additional discussions on the analysis procedure are found in SI.

\newpage
\section{Results}
\subsubsection{Keto Isocytosine}

Keto isocytosine, as a tautomer of cytosine (a fundamental DNA base), represents an important cyclic organic compound. Its photo-induced processes have attracted substantial attention in previous theoretical studies,~\cite{Nakayama,Mai,Gonzalez,Keto1,Keto2,Zhu2022,C6CP01391K} providing insights into the photochemical properties and non-adiabatic kinetic processes of nucleic acid bases.

Following TSH simulations, we collected all hopping geometries and initiated the analysis protocol. 
Previous studies indicated that hydrogen atoms in keto isocytosine do not significantly influence the nonadiabatic decay pathway, 
while their large-amplitude motions may lead to overestimation of their contribution to nonadiabatic dynamics. 
Therefore, we initially excluded all hydrogen atoms to focus primarily on motions of the ring carbon skeleton. 
We also performed parallel analyses including hydrogen atoms; however, these results proved less satisfactory and are documented solely in SI. 
We pursued both analytical approaches because ML methods are typically employed in a black-box manner, 
making it difficult to ascertain a priori whether hydrogen atoms play essential roles in the nonadiabatic dynamics. 
Thus it is not trivial to determine whether their contributions should be neglected in result analyses. 
In practical applications, we generally recommend testing both approaches and selecting the one yielding superior results.

Six sets of molecular descriptors were generated. 
For high-dimensional descriptors (MBTR, SOAP, and AEV), PCA served as a preprocessing step to reduce dimensionality. Components capturing over 99\% of the total variance were retained for the subsequent dimension reduction. As demonstrated in Table \ref{tab:ketopca}, this approach achieved significant dimensionality compression while preserving critical information.

However, even after PCA processing, direct visualization and analysis of data distributions in these reduced spaces remained challenging. 
As shown in  Table \ref{tab:ketopca}, the variance contributions of the first two principal components for each descriptor set remained relatively small. 
This indicates that utilizing only the first two principal components for dimensionality reduction is inadequate, 
as their cumulative contributions remain insufficient due to the inherently nonlinear nature of the data distributions. 
Representing the original data without significant information loss would still require higher dimensionality even after PCA. 
As it is rather challenging to understand the high dimensional data for human being, further dimensionality reduction via Autoencoder is necessary.

\begin{table}[ht!]
\caption{PCA results for three high-dimensional molecular descriptors (keto isocytosine).}
\label{tab:ketopca}
\begin{tabular}{
>{\columncolor[HTML]{FFFFFF}}c 
>{\columncolor[HTML]{FFFFFF}}c 
>{\columncolor[HTML]{FFFFFF}}c 
>{\columncolor[HTML]{FFFFFF}}c 
>{\columncolor[HTML]{FFFFFF}}c }
\hline
\begin{tabular}[c]{@{}c@{}}Molecule\\ Descriptor\end{tabular} & \begin{tabular}[c]{@{}c@{}}Dimension of\\ Initial Features\end{tabular} & \begin{tabular}[c]{@{}c@{}}Numbers of  Features\\ after PCA\\ (variance ratio \textgreater 0.99)\end{tabular} & \begin{tabular}[c]{@{}c@{}}variance ratio\\ of RC1\end{tabular} & \begin{tabular}[c]{@{}c@{}}variance ratio\\ of RC2\end{tabular} \\ \hline
MBTR                                                          & 1020                                                                  & 51                                                                                                            & 0.272                                                           & 0.137                                                           \\
SOAP                                                          & 6447                                                                  & 17                                                                                                            & 0.369                                                           & 0.147                                                           \\
AEV                                                           & 7596                                                                  & 32                                                                                                            & 0.387                                                           & 0.141                                                           \\ \hline
\end{tabular}
\end{table}

\begin{figure}[ht]
    \centering
    \includegraphics[width=0.9\linewidth]{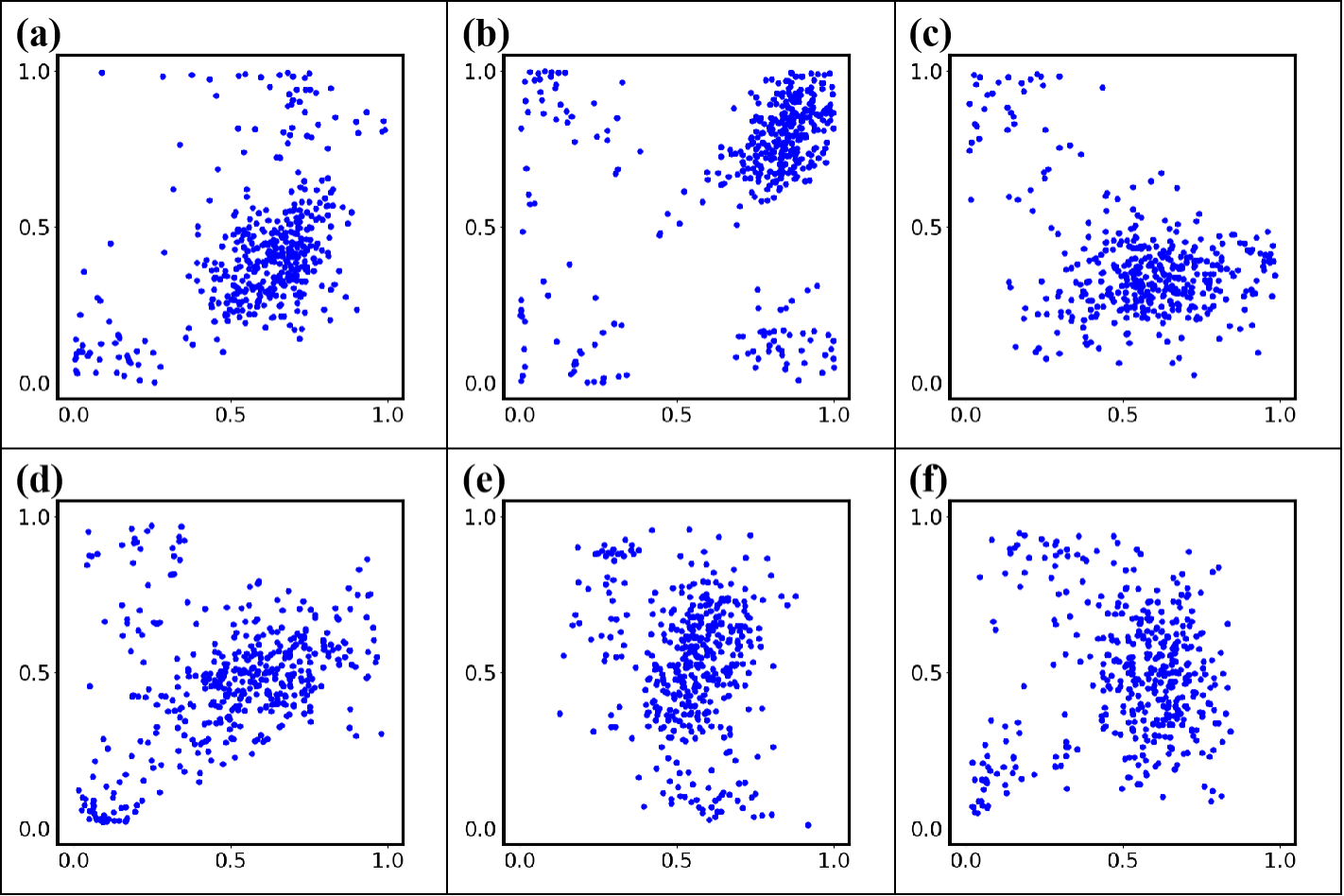}
    \caption{
    Dimensionality reduction results of descriptors via the Autoencoder for Keto isocytosine model. Different descriptors are employed: (a) Cartesian coordinates; (b) RIC; (c) IDM; (d) MBTR; (e) SOAP; (f) AEV.
    }
    \label{fig:fig3}
\end{figure}

The dimensionality reduction results for hopping geometries after Autoencoder compression are presented in Figure~\ref{fig:fig3}. 
In principle, a transparent analysis for multi-channel reactions should satisfy the following criteria: 
First, all data should separate clearly into several groups, each representing a distinct reaction channel. 
Second, cluster boundaries should be well-defined in the two-dimensional reduced space for clear visualization. 
Third, the emergence of more clusters in the two-dimensional space generally indicates better results, 
as it suggests identification of more distinct channels.

According to these criteria, RIC emerges as the optimal descriptor set for the current system. 
As shown in Figure~\ref{fig:fig3}(b), several high-density clusters form with clear separation by low-density regions. 
Four clusters are distinctly identified, along with unambiguous recognition of noise points.

For Cartesian coordinates and IDM, dimensionality reduction via Autoencoder also successfully assigns hopping geometries into a few groups with reasonably clear boundaries, with subsequent clustering analysis potentially identifying two and three clusters, respectively. 
Thus, both descriptors remain viable for further analyses. 
However, when employing RIC, four clusters are unambiguously generated. 
Clearly, identifying more clusters reduces the probability of missing reaction channels. 
Therefore, RIC descriptors undoubtedly represent the superior choice here. 
MBTR descriptors also appear to recognize three reaction channels, but their boundaries lack clarity, complicating subsequent clustering analysis. 
The final two descriptors, SOAP and AEV, yield unsatisfactory results, 
making different reaction channels difficult to identify in the reduced two-dimensional space.

Overall, RIC provides the best performance among all descriptors. 
The underlying reason may be explained as follows: Ring deformations play crucial roles in the nonadiabatic dynamics of keto-isocytosine, governed primarily by torsional motions of the chemical bonds within the ring moiety. 
Among all six descriptors, only RIC explicitly incorporates information about molecular dihedral angles.

\begin{figure}[ht]
    \centering
    \includegraphics[width=0.9\linewidth]{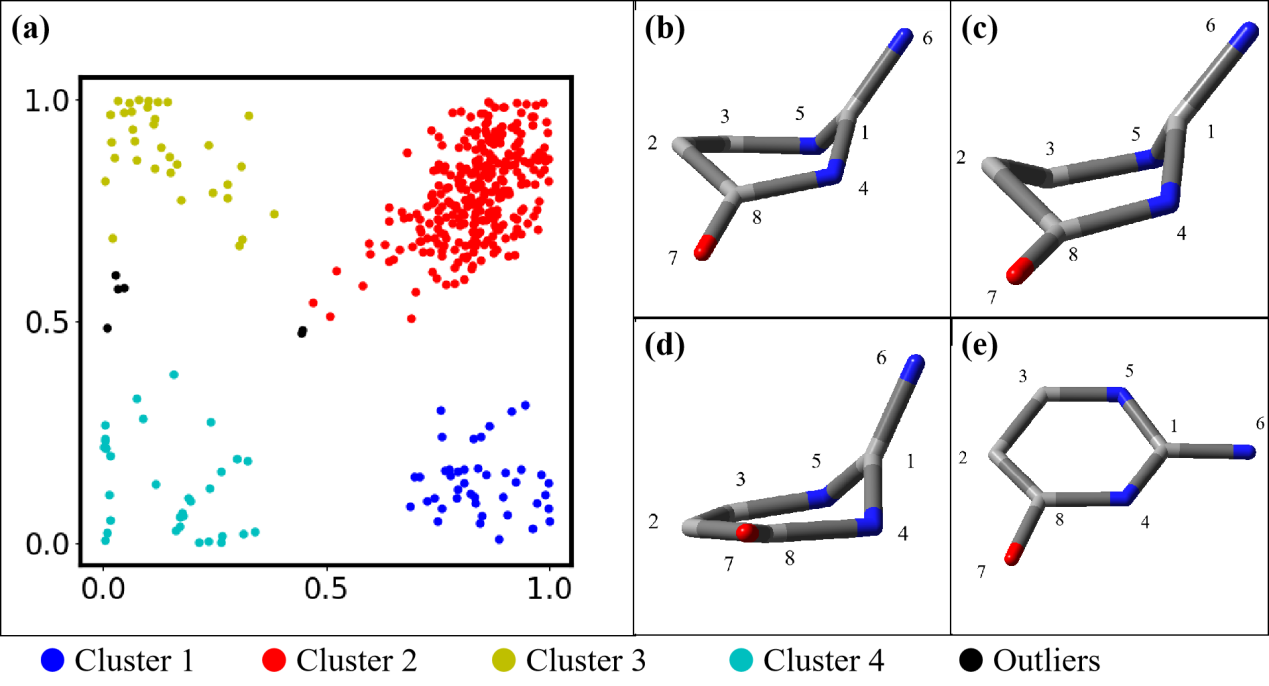}
    \caption{
    (a) DBSCAN results of RIC descriptors ($\varepsilon$ = 0.15 and $\mathrm{MinPts}$ = 8). (b)-(e) The typical molecular structures of Cluster 1-4.
    }
    \label{fig:fig4}
\end{figure}

We subsequently selected the dimensionality reduction results based on RIC descriptors for further analysis. As depicted in Figure~\ref{fig:fig4}(a), four clusters are successfully identified through clustering analysis based on data points in the reduced space. The four clusters are labeled as Cluster 1-4, which contain the $10.1\%$, $74.0\%$, $8.2\%$, $7.7\%$ of all molecules respectively. The ratio are calculated after removing the noise points.
Within appropriate parameter ranges, DBSCAN results show minimal dependence on parameter selections, and cluster profiles remain stable across different parameter setups. This approach consistently yields four clusters with clear boundaries and limited noise points as outliers. 
In principle, each individual non-separable cluster corresponds to a distinct reaction channel. To obtain preliminary insights into the geometric characteristics of each decay channel, we selected the center point of each cluster as the representative structure for the corresponding decay channel, as shown in the Figure~\ref{fig:fig4}(b)-(e).  

Among the four groups, three (Clusters 1, 2, and 3) share substantial geometric similarities, 
all exhibiting strong out-of-plane ring deformation at the C1 atom site. 
Their individual geometric features are summarized as follows: Both Cluster 1 and 2 display out-of-plane motion at the C2 site as well, 
differing primarily in the orientation of the NH$_2$ group, characterized by the pyramidalization status at the C1 atom site. 
Additionally, Cluster 1 exhibits C8 ring puckering motion. 
Cluster 3 shows no additional ring deformation sites beyond the C1 atom. 
Cluster 4 presents a distinctly different profile, maintaining a planar structure with observable CO bond stretching.

Although visual inspection provides initial assessment of hopping structure characteristics, 
identifying all essential reactive coordinates remains challenging due to the involvement of numerous internal coordinates. 
Therefore, more comprehensive analysis is necessary to determine which molecular motions play essential roles in characterizing each nonadiabatic decay channel. 
While establishing transformation relationships between reduced and original coordinates within the Autoencoder framework is theoretically possible, 
constructing this correspondence proves difficult due to the underlying complex nonlinear expressions.
Therefore, we utilized information entropy to analyze correlations between clustering results and molecular motions. 

Lower entropy values indicate more important features. 
Since entropy values range between 0 and 1, 0.5 represents a natural boundary between important and non-important features. 
For refined analysis, we calculate the distribution of all conditional entropy values and select the minimum point near 0.5 as the definitive boundary.

\begin{figure}[ht]
    \centering
    \includegraphics[width=0.9\linewidth]{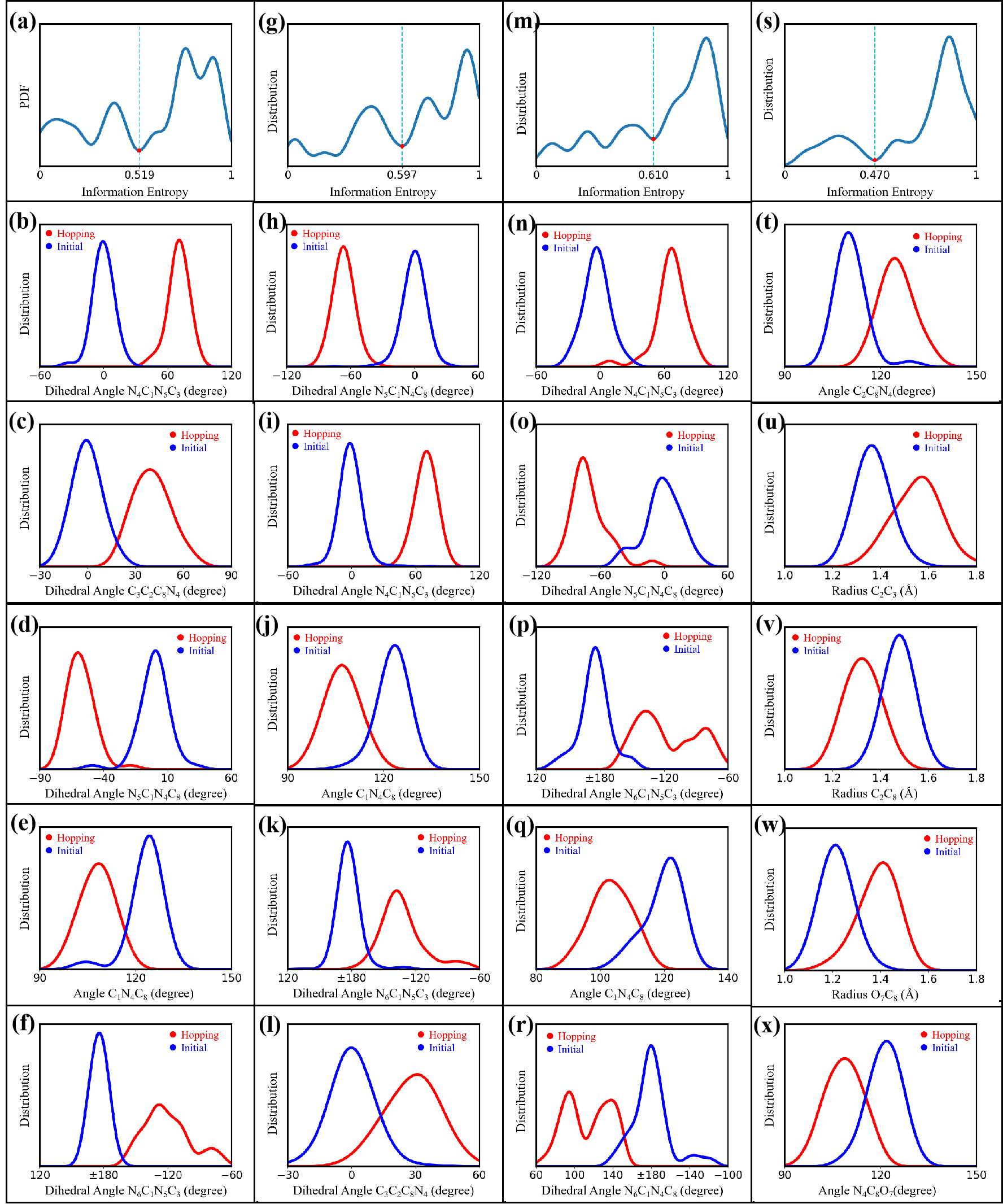}
    \caption{
    (a),(g),(m),(s) The distribution of information entropy of Cluster 1-4. (b)-(f) The distribution of D(4,1,5,3), D(3,2,8,4), D(5,1,4,8), A(1,4,8) and D(6,1,5,3) of Cluster 1 respectively. (h)-(l) The distribution of D(5,1,4,8), D(4,1,5,3), A(1,4,8), D(6,1,5,3) and D(3,2,8,4) of Cluster 2 respectively. (n)-(r) The distribution of D(4,1,5,3), D(5,1,4,8), D(6,1,5,3), A(1,4,8) and D(6,1,4,8) of Cluster 3 respectively. (s)-(x) The distribution of A(2,8,4), R(2,3), R(2,8), R(7,8) and A(4,8,7) of Cluster 4 respectively.
    }
    \label{fig:fig5}
\end{figure}

As shown in Figure~\ref{fig:fig5}, we present analysis results for the four decay channels, including the distribution of information entropy and the five most important features for each cluster. 
Due to space constraints, all additioal results are provided in the SI. 
For reference, all dihedral angles in initial structures 
are distributed close to 0 or $\pm$180 degrees, characteristic of near-planar configurations.

\clearpage

Among the four identified clusters, we first focus on Cluster 2, which contains the $74.0\%$ of molecules. 
Information entropy analysis results are presented in  Figure~\ref{fig:fig5}(g)-(l). 
Based on the information entropy distribution in Figure~\ref{fig:fig5}(g), 
we select 0.597 as the threshold value to separate important and unimportant geometric features according to the previously discussed principle. 
Eleven important features are identified, sorted by ascending entropy values: D(5,1,4,8), D(4,1,5,3), A(1,4,8), D(6,1,5,3), D(3,2,8,4), R(1,5), D(6,1,4,8), A(1,5,3), D(8,2,3,5), A(4,1,5) and D(2,3,5,1). 
The distributions of the top five internal coordinates are shown in Figure~\ref{fig:fig5}(h)-(l), others are shown in Figure~S2(g)-(l). 
Since these internal coordinate elements with small information entropy exhibit significant differences between initial and hopping structures,
we confirm that our analysis correctly identifies key features governing the nonadiabatic dynamics.

We now provide further analysis of the reactive coordinates for the nonadiabatic decay channel corresponding to Cluster 2. 
The first two components, D(5,1,4,8) and D(4,1,5,3), show significant changes of approximately 70 degrees from initial to hopping geometries.
Five other dihedral angles, D(6,1,5,3), D(3,2,8,4), D(6,1,4,8), D(8,2,3,5), and D(2,3,5,1), also exhibit noticeable changes of 20-40 degrees. 
These findings align with the representative hopping geometry of this cluster. 
For instance, the pronounced ring deformation at the C1 atom site is well characterized by the two leading components D(5,1,4,8) and D(4,1,5,3), 
along with other dihedral angles involving C1 [D(6,1,5,3), D(6,1,4,8), and D(2,3,5,1)]. 
Simultaneously, distortion at the C2 atom site produces visible changes in dihedral angles D(3,2,8,4), D(8,2,3,5), and D(2,3,5,1). 
More precisely, trajectories in Cluster 2 undergo conformational transformation from planar geometry at the ground state minimum to a boat-like structure at the conical intersection. 
This structural transition can be explained by evolution of the $\pi$-system during nonadiabatic dynamics: 
photoexcitation breaks delocalized $\pi$ bonds, 
leading to adoption of a non-planar boat conformation to minimize steric hindrance, 
accompanied by substantial modifications of bond lengths and angles within the ring moiety. 
Importantly, these geometric evolutions are effectively captured by the specific internal coordinates selected through information entropy analysis based on RIC dimensionality reduction.


Analysis results for the decay channel corresponding to Cluster 1 are shown in Figure~\ref{fig:fig5}(a)-(f) and Figure~S2(a)-(f). 
Following similar criteria, we select 0.519 as the threshold, 
identifying 11 important features in descending order: D(4,1,5,3), D(3,2,8,4), D(5,1,4,8), A(1,4,8), D(6,1,5,3), D(3,2,8,7), D(6,1,4,8), D(8,2,3,5), A(1,5,3), R(1,5), D(2,3,5,1). 
Several features play essential roles in both Cluster 1 and 2, as both exhibit ring deformation at the C1 atom site. 
However, different rankings and contributions of involved features indicate geometric distinctions between these clusters. 
For example, increased contributions of D(3,2,8,4) and D(3,2,8,7) in Cluster 1 reflect the involvement of C8-puckering motion that drives the ring moiety away from a simple boat conformation.

As shown in Figures~\ref{fig:fig5}(n)-(r) and Figure~S2(m)-(q), 
we select 0.610 as the threshold for information entropy analysis of Cluster 3. 
Ten important features are identified: D(4,1,5,3), D(5,1,4,8), D(6,1,5,3), A(1,4,8), D(6,1,4,8), A(4,1,5), D(1,4,8,2), D(2,3,5,1), D(1,4,8,7) and R(1,5). 
Numerous features appear in previous analyses of Cluster 1 and 2 due to geometric similarities. 
However, certain features, such as D(3,2,8,4), A(1,5,3) and D(8,2,3,5), play more important roles here, 
accompanied by declining contributions from D(1,4,8,2) and D(1,4,8,7). 
The underlying reason is clear: ring deformation is localized primarily at the C1 site, while the remainder of the six-membered ring maintains a planar structure.

These analyses reveal that Clusters 1-3 share substantial geometric similarities, 
particularly the presence of C1-puckering motion. 
Their distinctions arise from varying puckering motions at C2 and C8 sites, along with differing pyramidalization at the C1 site. 
This suggests that Clusters 1-3 may reside along the same conical intersection seam. 
While constructing such a CI seam would provide more comprehensive understanding, this endeavor lies beyond the scope of the current work. 
In this study, we prioritize identifying more reaction channels and automating the identification of leading reaction coordinates for each channel. 
From this perspective, enhancing the efficiency of the automated working protocol without losing important reaction channels represents a more critical objective.

For Cluster 4, we select 0.470 as the threshold, identifying only 6 important features: A(2,8,4), R(2,3), R(2,8), R(7,8), A(4,8,7), and A(1,4,8), 
as shown in Figure~\ref{fig:fig5}(t)-(x) and Figure~S2r.  
Remarkably, no dihedral angles appear among the important features due to the planar structure of Cluster 4. 
In this situation, $\pi$ electrons delocalize across the entire C3-C2-C8-O7 framework, rather than localizing in individual double bonds (C3-C2 and C8-O7) as in initial configurations. This delocalization causes three bond lengths to be similar, reflected by the inclusion of these bond lengths in the information entropy analysis.

These findings align with previous studies.~\cite{Keto2,Zhu2022} 
In one of the previous work~\cite{Zhu2022},  a hierarchical analysis protocol was introduced to systematically investigate the nonadiabatic dynamics of this system. Within the iterative framework, RICs were employed as descriptors and categorized into six distinct sets. PCA and clustering analyses were subsequently performed three times using one of these descriptor sets. Initially, all hopping geometries were separated into two clusters, while the larger cluster can be divided again. In the second iteration, this major cluster was resolved into three well-defined reaction channels, resulting in a total of four clusters—three corresponding to clear reaction pathways. In the final iteration, one cluster was again divided into three subchannels, ultimately identifying six reaction channels.
In comparison, our current framework streamlines the analysis process by eliminating the need for manual partitioning of RICs. We simply apply Autoencoder-based dimensionality reduction to the complete set of RICs in a single step, followed by information entropy analysis to identify active coordinates within a unified workflow. The present protocol identifies four channels, consistent with the results obtained after the second iteration of the previous approach. Although the current method does not yield all six channels identified in the final iteration of the earlier study, the subtle geometrical differences distinguishing the three subchannels obtained in the third iteration in the previous work do not significantly impact the analysis of nonadiabatic dynamics. Given its simpler one-step nature, the current protocol offers an excellent balance between computational accuracy and efficiency in characterizing nonadiabatic reaction dynamics. Moreover, the optimized conical intersection geometries obtained show excellent agreement with those reported in earlier work using manual assignment~\cite{Keto2}, confirming that our approach provides an efficient and straightforward framework for analyzing nonadiabatic reaction mechanisms.

\newpage
\subsection{Methaniminium Cation}

The methaniminium cation,~\cite{CH4N1,Lixusong1,Newton-X1,Fabiano,codeTSH,Tapavicza,Barbatti} one of the simplest organic nitrogen-containing cations, is widely employed as a prototype for studying structure-property relationships in bio-relevant compounds such as retinal. Consequently, considerable theoretical studies have explored its photochemistry, 
both to clarify photo-reaction mechanisms in similar systems and to benchmark newly developed theoretical methods. In this work, we also extracted reaction channels and relevant key coordinates from TSH simulations of nonadiabatic dynamics of  methaniminium cation. Following the detailed working protocol described in previous sections and SI, we conducted TSH dynamics starting from the second excited state $S_2$, collected $S_1 \longrightarrow S_0$ hopping geometries, converted them into different molecular descriptors, performed initial data preprocessing, conducted dimensionality reduction via Autoencoder, make clustering analysis to identify reaction channels, and finally applied information entropy to extract key coordinates for each channel.

The results of preprocessing of MBTR, SOAP, and AEV descriptors shows that PCA significantly reduces input data dimensionality, 
as shown in Table 3. Due to nonlinear patterns in data distribution, 
using only the two largest principal components from PCA for further analysis proves inadequate, necessitating the Autoencoder approach.

\begin{table}[]
\caption{PCA results for three high-dimensional molecular descriptors (methaniminium cation)}
\label{tab:cationpca}
\begin{tabular}{
>{\columncolor[HTML]{FFFFFF}}c 
>{\columncolor[HTML]{FFFFFF}}c 
>{\columncolor[HTML]{FFFFFF}}c 
>{\columncolor[HTML]{FFFFFF}}c 
>{\columncolor[HTML]{FFFFFF}}c }
\hline
\begin{tabular}[c]{@{}c@{}}Molecule\\ Descriptor\end{tabular} & \begin{tabular}[c]{@{}c@{}}Dimension of\\ Initial Features\end{tabular} & \begin{tabular}[c]{@{}c@{}}Numbers of  Features\\ after PCA\\ (variance ratio \textgreater 0.99)\end{tabular} & \begin{tabular}[c]{@{}c@{}}variance ratio\\ of RC1\end{tabular} & \begin{tabular}[c]{@{}c@{}}variance ratio\\ of RC2\end{tabular} \\ \hline
MBTR                                                          & 702                                                                  & 62                                                                                                            & 0.247                                                           & 0.102                                                           \\
SOAP                                                          & 4344                                                                  & 20                                                                                                            & 0.541                                                           & 0.101                                                           \\
AEV                                                           & 4080                                                                  & 54                                                                                                            & 0.319                                                           & 0.128                                                           \\ \hline
\end{tabular}
\end{table}

\begin{figure}[ht]
    \centering
    \includegraphics[width=0.9\linewidth]{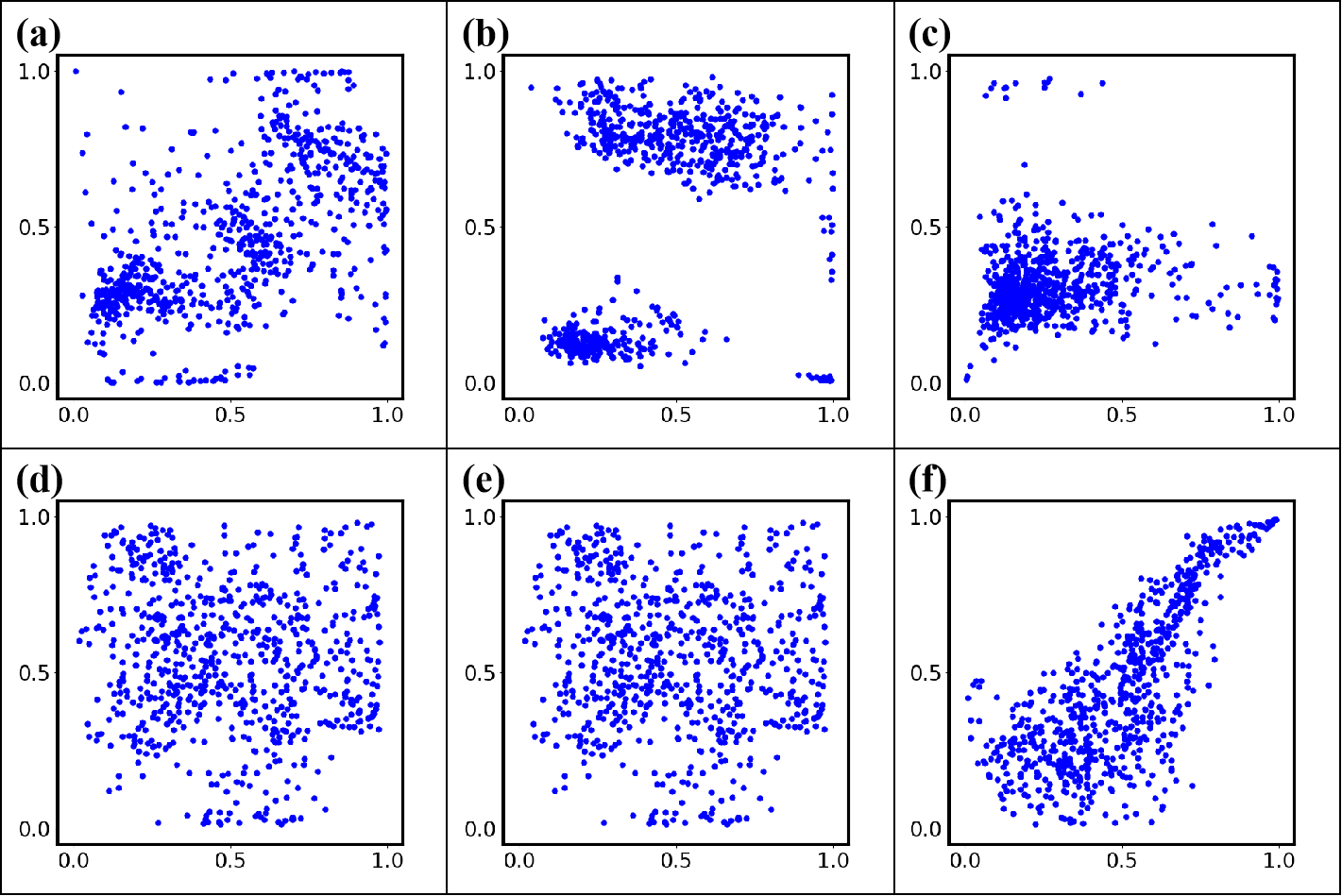}
    \caption{
    Dimensionality reduction results of descriptors via the Autoencoder for methaniminium cation model. Different descriptors are employed: (a) Cartesian coordinates; (b) RIC; (c) IDM; (d) MBTR; (e) SOAP; (f) AEV.
    }
    \label{fig:fig6}
\end{figure}

Autoencoder dimensionality reduction results are shown in Figure~\ref{fig:fig6}.  
When using MBTR, SOAP, and AEV descriptors, no obvious separation of data points into distinct groups emerges. 
In contrast, Cartesian coordinates and IDM descriptors show potential for separating a small subset of points from high-density regions, 
though determining whether these represent meaningful clusters or noise remains challenging. 
Only with RIC descriptors are hopping geometries clearly divided into a few of categories. 
This strongly suggests that RIC should be used for subsequent analyses.

\begin{figure}[ht]
    \centering
    \includegraphics[width=0.9\linewidth]{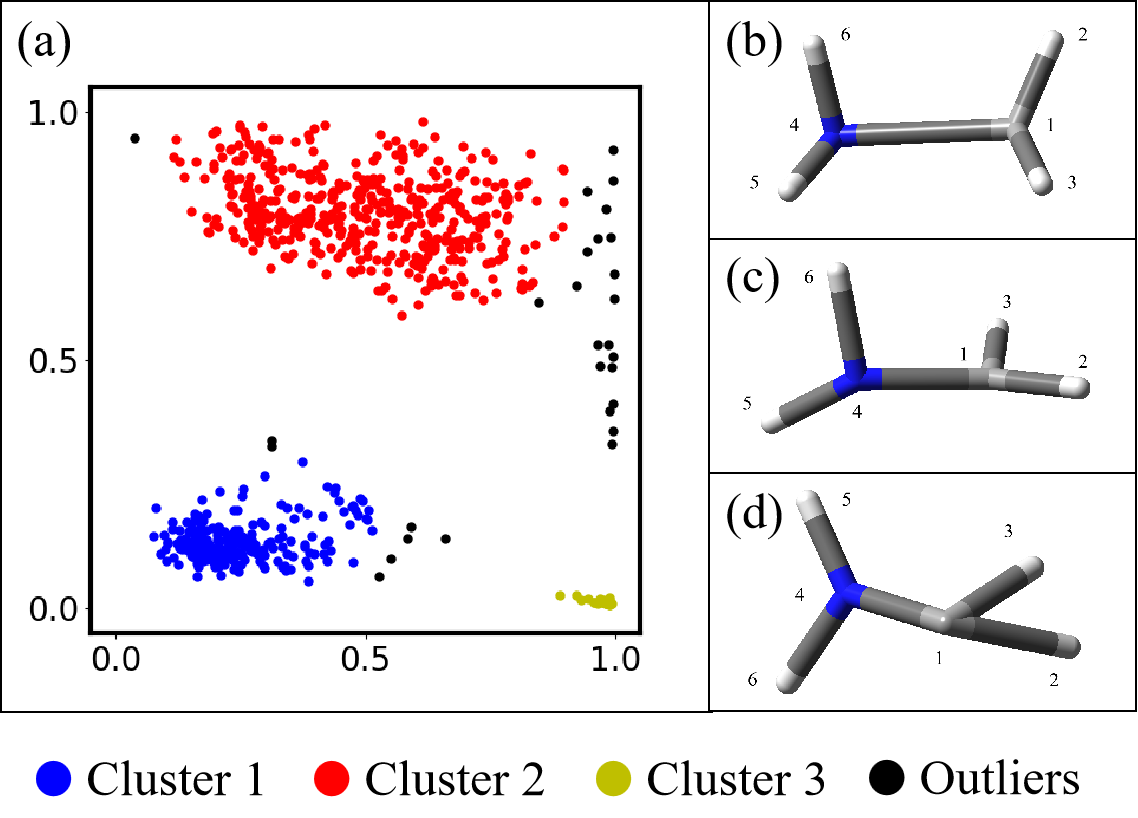}
    \caption{
    (a) DBSCAN results of RIC descriptors ($\varepsilon$ = 0.05 and $\mathrm{MinPts}$ = 15). (b)-(d) The typical molecular structures of Cluster 1-3.
    }
    \label{fig:fig7}
\end{figure}

Clustering analysis of RIC-based dimensionality reduction results is presented in Figure~\ref{fig:fig7}(a). 
Three clusters are identified, with a limited number of data points designated as noise. The three clusters are labeled as Cluster 1-3, which contain the $35.0\%$, $60.0\%$, $5.0\%$ of all molecules respectively. The ratio are calculated after removing the noise points.
Specifically, we treat all data in the lower-right region as a separate cluster. 
In test calculations, these points could be assigned as an individual cluster or treated as noise, depending on the MinPts value. 
However, to avoid missing possible reaction channels, we treat them as meaningful data forming a distinct cluster. 
This division proves reasonable, as the three channels ultimately exhibit distinct reaction coordinates.

Representative structures for each cluster are shown in Figure~\ref{fig:fig7}(b)-(d), respectively. These structures illustrate the characteristic geometric features of each cluster: Cluster 1 displays the significant CN bond elongation and the bi-pyramidalization at both C and N sites; Cluster 2 exhibits the strong CN torsion with nearly perpendicular carbon and nitrogen planes; Cluster 3 shows the significantly pyramidalization at the C atom, driven by the out-of-plane motions of two H atoms in the CH$_2$ group.

\newpage

\begin{figure}[ht]
    \centering
    \includegraphics[width=0.7\linewidth]{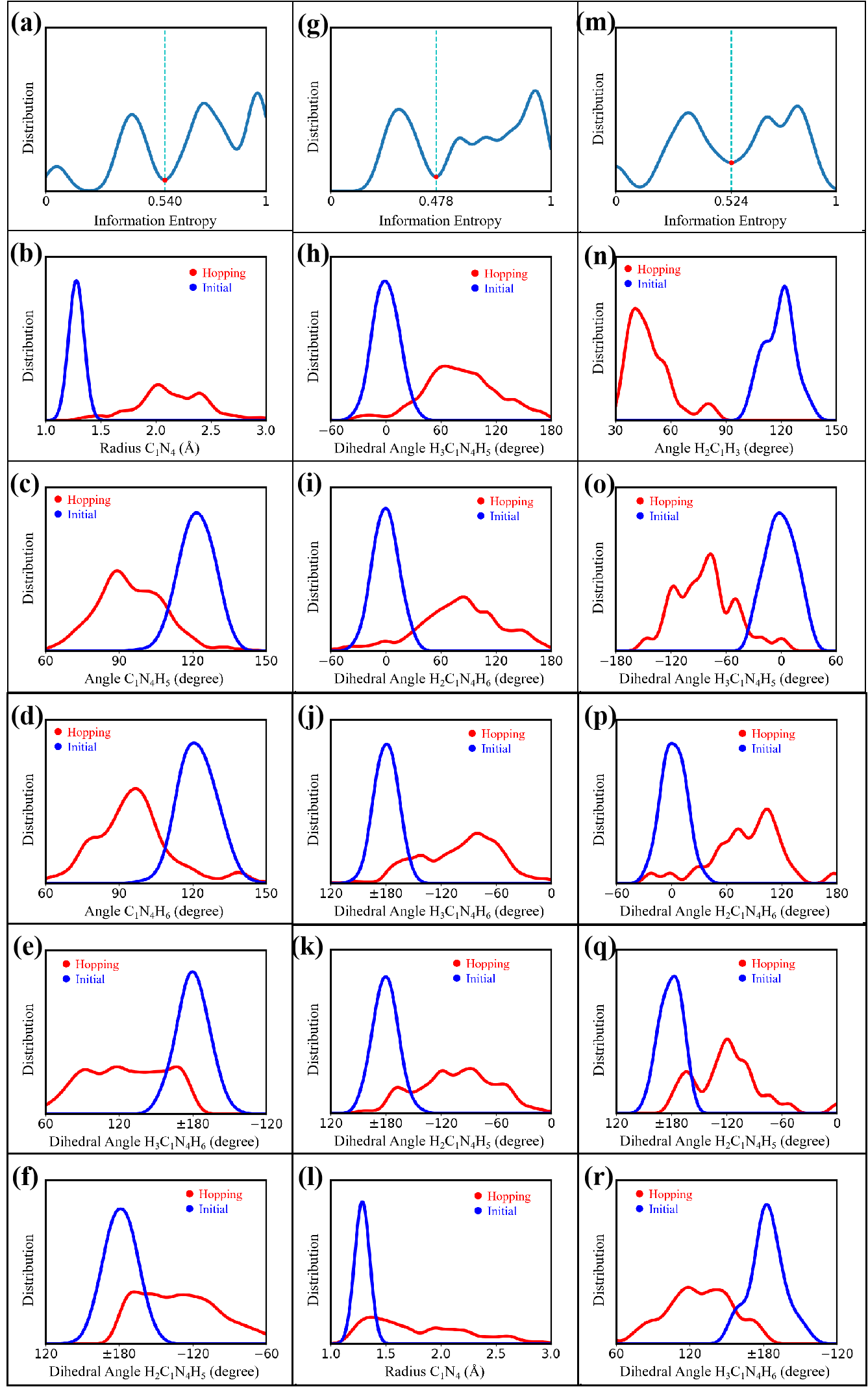}
    \caption{
    (a),(g),(m) The distribution of information entropy of Cluster 1-3. (b)-(f) The distribution of R(1,4), A(1,4,5), A(1,4,6), D(3,1,4,6) and D(2,1,4,5) of Cluster 1 respectively. (h)-(l) The distribution of D(3,1,4,5), D(2,1,4,6), D(3,1,4,6), D(2,1,4,5) and R(1,4) of Cluster 2 respectively. (n)-(r) The distribution of A(2,1,3), D(3,1,4,5), D(2,1,4,6), D(2,1,4,5) and D(3,1,4,6) of Cluster 3 respectively.
    }
    \label{fig:fig8}
\end{figure}

Information entropy analysis was also performed for methaniminium cation, as summarized in Figure~\ref{fig:fig8} and SI.

Following the previously established criterion, we select 0.540 as the boundary, identifying R(1,4), A(1,4,5), A(1,4,6), D(2,1,4,5) and D(3,1,4,6) as important features for Cluster 1. The entropy of R(1,4) is significantly lower than other features, indicating the dominant role of the C1-N4 bond elongation as the primary feature of Cluster 1  (Figure~\ref{fig:fig8}(b)). The other important features, including A(1,4,5) and A(1,4,6), D(2,1,4,5) and D(3,1,4,6), indicating the pyramidalization motion at both C1 and N4 atoms. According their entropy values, the degree of pyramidalization at C1 is less pronounced than at N4. 


For Cluster 2 molecules, D(3,1,4,5), D(2,1,4,6), D(3,1,4,6), D(2,1,4,5) and R(1,4) constitute important features, with information entropy values below 0.478. The most distinctive features of Cluster 2 originate from dihedral angle changes caused by the C1-N4 torsion, with all four dihedral angles changed (Figure~\ref{fig:fig8}(h)-(k)). The C1-N4 bond elongation is less pronounced than in Cluster 1.


For Cluster 3 molecules, we select 0.524 as the threshold, identifying 7 features with lower entropy: A(2,1,3), D(3,1,4,5), D(2,1,4,6), D(2,1,4,5), D(3,1,4,6), R(1,2) and R(1,3). The information entropy value for A(2,1,3) reaches the theoretical minimum, as shown in Figure~\ref{fig:fig8}(n), the two hydrogen atoms bonded to C1 show the out-of plane motion, forming the pyramidalization at the C1 site and driving two H atoms closer. 
Also due to the motion of H2 and H3 atoms, the bond lengths R(1,2) and R(1,3) become important features (Figure~S2(s)-(t)). 

The methaniminium cation~\cite{CH4N1,Lixusong1,Newton-X1,Fabiano,codeTSH,Tapavicza,Barbatti} is typical model system which has been studied by many previous work. All these work points out there are two decay channels (labeled I and II) in this molecule which the trajectories start from $S_2$. The group I characterized by significant CN bond elongation accompanied by C/N atom bipyramidalization, corresponds to our Cluster 1. The group II, featuring weak CN bond vibration mixed with CN torsion motion, aligns with our Cluster 2. Cluster 3, possibly due to its smaller population and the lack of key angle analysis, was not mentioned previously. This consistency with previous work demonstrates that our method achieves consistent and improving result results through a more automated procedure.


\clearpage

\newpage

\subsection{Discussions}

The current analysis protocol integrates several key computational algorithms into a powerful and efficient automated framework that can be readily extended to more general applications. We now discuss several crucial aspects of this framework.


It is certainly possible to perform the analyses based on other dimensionality reduction methods,~\cite{PCA1,PCA2,MDS1,MDS2,ISOMAP1,ISOMAP2,diffusionmap1,diffusionmap2} like PCA~\cite{PCA1,PCA2} and MDS.~\cite{MDS1,MDS2}
Compared with them, the Autoencoder show several obvious advantages. 
First, the Autoencoder shows the excellent performance to  processes both linear and nonlinear data distribution. This strength stems from their use of nonlinear activation functions in hidden layers, which break the constraints of linearity and allow the network to learn complex relationships within data. In contrast, the standard PCA and MDS methods are linear approaches essentially. 
Through some variants, like kernel PCA or others, they can handle mild nonlinearity by predefining nonlinear mappings, while these models still lack the flexibility to learn nonlinearities directly from the data itself. 
Second, for the large datasets with the extremely high data dimensions, PCA and MDS need to perform the diagonalization of a large matrix, which become computationally expensive and memory-heavy. As contrast, the Autoencoder allow the mini-bash trainings  that provide much better performance in computational costs when facing the large datasets.

However, the advantage of the Autoencoder also brings some 
shortcomings, such as the existence of the double sides in a coin.
Because of the very complicated nonlinear operation in the Autoencoder, it is not trivial to directly 
construct the direct transformation between the original and reduced coordinates in the Autoencoder, bringing the difficulties 
to identify which element of molecular descriptors are dominant
in the dynamics evolution. To deal with this situation,
the employment of the information entropy is highly useful.
Our work shows that the information entropy
plays a vital role in the analysis of reaction coordinates, providing a powerful tool for identifying the important molecular motions.
This approach reduces reliance on human intuition and experience during feature extraction.

Notably, RIC demonstrated superior representational capacity compared to other molecular descriptors in these analyses. 
This shows that it is necessary to explicitly add the dihedral angle information of the molecule to the descriptor as such information are generally important in the accurate description of molecular configurations.
The performances of other descriptors are not satisfied enough, as some of them may not distinguish all reaction channels. 
In practices, we also suggest to try several available molecular descriptors in the analyses of realistic systems. 
In addition, we also recommend to include the H atom or not to conduct two sets of parallel analyses.
As the current workflow is highly automated, it is more transparent to judge and choose the suitable descriptors during the analysis protocol via the doing-by-learning manner.  
We also expected that the consideration of multiple molecular descriptors promotes broader applicability of the current approach.


\newpage
\section{Conclusions}

In this study, we have developed an automated computational framework that integrates unsupervised machine learning techniques to analyze non-adiabatic dynamics simulations. 
This protocol is composed of two steps, the first step is to identify how many reaction channels are involved in the nonadiabatic dynamics evolution, and the second one tries to clarify which molecular motion is responsible for each decay channel. 

After the TSH dynamics, we collected the hopping geometries, built various molecular descriptors and perform the dimensionl reduction Autoencoder. In the two-dimensional reduced space, the distribution of all hopping geometries is constructed. Then we may use the dimensionl reduction results based on the proper molecular descriptors to conduct the clustering analyses, determine the number of the reaction channels. 
For each channel, the information entropy with the chosen descriptors were adopted to clarify the corresponding reaction coordinates. This comprehensive pipeline enables complete automation from raw trajectory data to mechanistic insights of nonadiabatic processes.

The framework's effectiveness has been demonstrated through applications to two representative molecular systems, keto isocytosine and methaniminium cation. 
For keto isocytosine, it successfully identified four distinct non-adiabatic decay channels and clarify their own geometrical features with different ring deformation patterns. For methaniminium cation, it revealed three reaction channels involving CN bond stretching, torsion, and pyramidalization.  Notably, RIC consistently demonstrated superior performance in capturing essential geometric features that distinguish different reaction pathways. The successful application to these chemically distinct systems confirms the robustness and general applicability of the current framework in non-adiabatic dynamics studies.

We wish to point out that it is very important to set up a highly automatic analysis protocol for more wide applications. In this sense, for more general problems, the development of easy-to-use black-box analysis package is necessary.~\cite{ULaMDyn} 
Along this idea, the integration the current and other available approaches, based on differnet molecular descriptors and ML algorithms, should be highly useful in the further.


This automated framework addresses fundamental challenges in traditional analysis of on-the-fly non-adiabatic molecular dynamics simulations, where the complexity of structural transformations, high dimensionality of trajectory data, and nonlinear nature of configuration space evolution present significant obstacles to conventional approaches. By integrating the constructions different molecular descriptors, Autoencoder-based nonlinear dimensionality reduction, clustering analysis and information entropy, our method automatically identifies photochemical reaction channels and their characteristic coordinates without requiring pre-knowledge or extensive manual interpretation. 
This approach substantially enhances the objectivity, efficiency, and reproducibility in analyzing complex non-adiabatic processes, establishing a solid foundation for automated interpretation of photochemical dynamics and opening new possibilities for systematic investigation of excited-state reaction mechanisms of more complex systems.


\section*{Supporting Information Available}

Several relevant information: More details of the theoretical methods and computational details, including TSH dynamics simulation, construction and preprocess of descriptors, PCA, Autoencoder; more additional dimensionality reduction results; information entropy analyze results; more distribution of important features; the structural information of the S0 minimum and typical one of each channel.

\section*{Author Information}
\subsection*{Corresponding Authors}
E-mail: zhenggang.lan@m.scnu.edu.cn

\subsection*{Notes}
The authors declare no competing financial interest.

\begin{acknowledgement}
    The authors express sincerely thanks to the National Natural Science Foundation of China (No. 22333003 and 22361132528) for financial support.
    Some calculations in this paper were done on SunRising-1 computing environment in Supercomputing Center, Computer Network Information Center, CAS.
\end{acknowledgement}

\newpage

\bibliography{reference}

\end{document}